\documentstyle[aps,prl,epsfig]{revtex}

\begin{document}

\title{Structural view of hexagonal non-perovskite AMnO$_3$}

\author{Bas B. van Aken$^*$\nocite{pal}, Auke Meetsma, and Thomas T. M. Palstra}
\address{Solid State Chemistry Laboratory, Materials Science Centre, University of Groningen,
\\ Nijenborgh 4, 9747 AG  Groningen, the Netherlands}

\date{\today}
\twocolumn[\hsize\textwidth\columnwidth\hsize\csname@twocolumnfalse\endcsname

\maketitle

\begin{abstract}
We refined the crystal structure of AMnO$_3$, with A = Y, Er, Yb and Lu by
single crystal x-ray diffraction. Our results show some distinct differences
with previous reports on LuMnO$_3$ and YbMnO$_3$. We show that the ferroelectric
behaviour is originated in the dipole moment at the A-site, and not, as is the
common opinion, at the Mn site.
\end{abstract}

\pacs{} ]

\section{Introduction}
In the search for new composition-properties relations ABO$_3$ compounds have
attracted a lot of attention. The perovskite materials, ABO$_3$, have been
researched extensively because this structure forms the basis for interesting
physical properties such as high $T_c$ superconductivity\cite{Bed86} and
colossal magnetoresistance\cite{Jin94a}. Non-perovskite AMnO$_3$, with A = Y,
Ho,...,Lu, attracted renewed interest, due to their ferroelectric
properties\cite{Ber63a}. These hexagonal AMnO$_3$\cite{Yak63} have a basically
different structure than most ABO$_3$ compounds, that are distorted perovskites.
These properties arise due to the strong correlation of the $3d$ electrons with
the O $2p$ orbitals. We will show in this paper that the ferroelectricity is
mostly due to the anomalous oxygen surrounding of the Lanthanide position. Note
that some perovskite ABO$_3$ compounds, like Sr-doped LaMnO$_3$ and LaCoO$_3$,
can have a rhombohedral-hexagonal structure. However, the hexagonal ABO$_3$ we
report here, have a profoundly different structure than the perovskite based
ABO$_3$ compounds.

This structure, including atomic positions, has been reported in literature
before. Yakel\cite{Yak63} initiated the research. Isobe et al. reported the
structure of YbMnO$_3$\cite{Iso91}. Mu\~{n}oz et al. used neutron powder
diffraction to study YMnO$_3$ and ScMnO$_3$\cite{mun00}. Furthermore
{\L}ukaszewicz reported the high temperature phase as P63/mmc\cite{Luk74}.
Remarkably the structural transition is some 300 K higher than the ferroelectric
ordering temperature. Note that we use O$_\textrm{X,ap}$ and O$_\textrm{X,eq}$
to denote apical and equatorial positions, respectively, with respect to the
cation X, with X = A or Mn. We observe that the crystal structure at ambient
temperature differs in several aspects from the previous reports, where the
local electric dipole moments were ascribed to the asymmetric Mn environment. We
show that it is caused by asymmetric coordination of the Y ion.

\section{crystal structure}\label{crystalstructure}

\begin{figure}[htb]
   \centering
   \includegraphics[height=80mm,width=67mm]{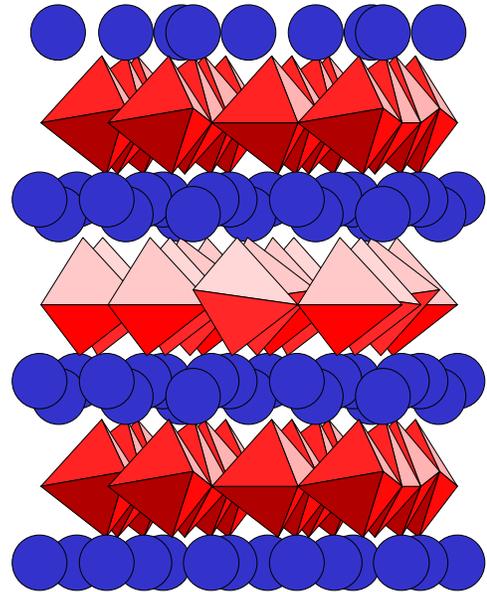}
   \caption{Sketch view along the basal plane of the crystal structure of AMnO$_3$. The A
cations are shown as circles, the MnO$_5$ are represented by trigonal
bipyramids. The figure highlights the two-dimensional nature of the crystal
structure.}
   \label{3d}
\end{figure}

\begin{figure}[htb]
   \centering
   \includegraphics[height=64mm,width=80mm]{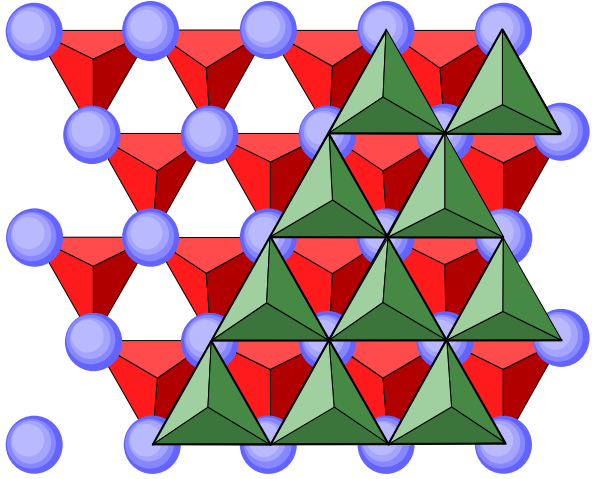}
   \caption{Sketch view along the $c$ axis of two layers, showing the complementary
occupation of half of the O$_5$ holes in each MnO layer. The Ln ions are located
exactly above the corner-linked O$_\textrm{Mn,eq}$ ions}
   \label{layer}
\end{figure}

To give an introduction to the hexagonal structure, we first take a look at the
centrosymmetric structure. The Mn ions are surrounded by a trigonal bipyramid of
oxygen ions. The bipyramids are linked by corner sharing the equatorial oxygens.
Between these slabs of bipyramids, there is a layer of A ions. In Fig.~\ref{3d}
the layered nature is shown.
The corner sharing is highlighted in Fig.~\ref{layer}. Each O$_\textrm{Mn,eq}$
is shared by three bipyramids. These oxygen ions make up a triangular lattice.
The lanthanide cations are located exactly between two O$_\textrm{Mn,eq}$ of
consecutive layers. In the left upper part of Fig.~\ref{layer}, we see that half
of the O$_\textrm{Mn,eq}$ triangles form trigonal bipyramids around Mn ions. In
the next layer, the other half is filled.

\section{electronic structure}\label{electronic structure}

\begin{figure}[htb]
   \centering
   \includegraphics[height=80mm,width=77mm]{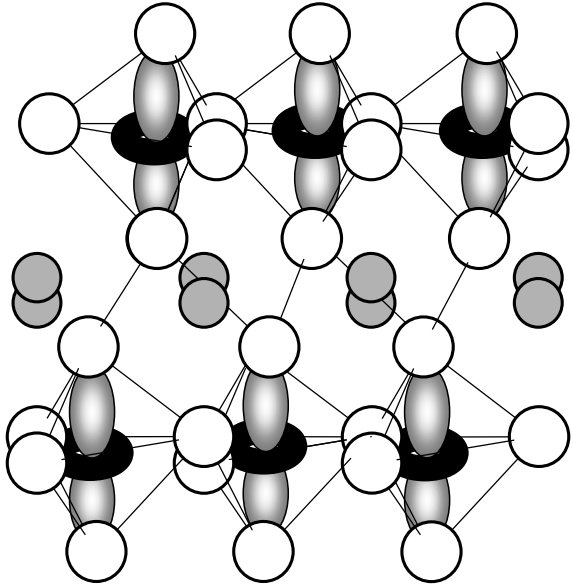}
   \caption{Sketch view of the Mn $3d$ $z^2$ orbitals in AMnO$_3$. Due to the crystal field
splitting of the trigonal bipyramid, the $z^2$ orbital is highest in energy and
in Mn$^{3+} 3d^4$ unoccupied. The figure shows also the very poor overlap
between two consecutive Mn $z^2$ orbitals.}
   \label{elect}
\end{figure}

We have calculated the crystal field splitting of a trigonal bipyramidal
field\cite{Van01b}. The Mn $3d$ levels are split according to their magnetic
quantum number. This yields the $xz, yz$ doublet with the lowest energy,
followed by the $xy, x^2-y^2$ doublet. The $3z^2-r^2$ orbital has the highest
energy. For YMnO$_3$, with Mn$^{3+} 3d^4$, the $z^2$ orbital is unoccupied. In
Fig.~\ref{elect}, the $z^2$ orbitals are shown for two layers.

\section{HT vs. LT}\label{HT vs. LT}

\begin{figure}[htb]
   \centering
   \includegraphics[height=57mm,width=81.5mm]{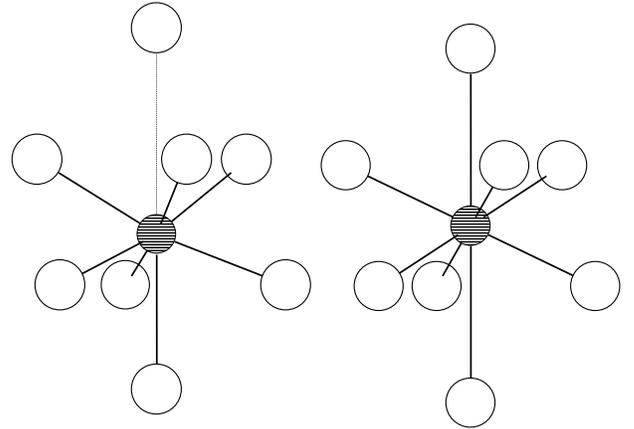}
   \caption{Sketch view of the surrounding oxygen polyhedra of the A ion. The left panel
shows the ferroelectric low temperature structure, the right panel the
centrosymmetric high temperature structure. The anomalous long bond is shown as
a thin line.}
   \label{LnO7}
\end{figure}

All hexagonal AMnO$_3$ are reported to be ferroelectric with transition
temperatures between 800 and 1000 K. For YMnO$_3$, the temperature dependence of
the structure is studied by observing the temperature dependence of the
integrated intensity of some reflections. The study showed that above the
transition temperature a structural transition to higher symmetry is observed.
The increase in symmetry yields a smaller unit cell volume, from Z=6 to Z=2. The
main difference in atomic positions between the high and low temperature
structure is that in the HT phase all atoms are constrained to planes, parallel
with the $ab$ plane. Below the transition temperature the structure loses the
mirror planes parallel to the $ab$ plane and all inequivalent atoms get a
refinable $z$ position. The deviations from these planes are sketched in
Fig.~\ref{LnO7}.
The deviations yield different bond lengths for A-O$_\textrm{A,ap}$. One bond
adopts a regular value $\sim2.4$ \AA\, while the other becomes about 1 \AA\
larger. The asymmetric A environment is the main reason for the ferroelectric
behaviour. As we have two inequivalent lanthanide positions in $P6_3cm$, we have
two inequivalent, although similar, dipole moments as a result of the movement
of the O$_\textrm{A,ap}$ ions. This movement can also be expressed as a tilting
of the MnO$_5$ bipyramids. Two out of three O$_\textrm{Mn,eq}$ positions in the
triangular base of the bipyramid are constraint by symmetry, note
O$_\textrm{A,ap}=$O$_\textrm{Mn,eq}$. The two equivalent O ions move down, the
other moves up. This yields, for the two layers in the unit cell, four upwards
pointing local dipole moments, whereas the other two point downwards.

\section{Data}

Single crystals of AMnO$_3$, A=Y, Yb, Er and Lu, were obtained using a flux
method by weighing appropriate amounts of A$_2$O$_3$ and MnO$_2$ with
Bi$_2$O$_3$ in a 1:12 ratio \cite{Yak63}. The powders were thoroughly mixed and
heated for 48 h at 1523 K in a Pt crucible. The separation of the crystals from
the flux has been done by increasing the temperature to 1723 K and evaporating
the Bi$_2$O$_3$ flux \cite{Ber63a}.

We have studied the crystal structure by single crystal x-ray diffraction.
Details of the measurement and the refinement are published elsewhere,
\emph{cf.} YMnO$_3$ \cite{Van01a}, ErMnO$_3$ \cite{Van01f}, YbMnO$_3$ and
YMnO$_3$ \cite{Van01e}. The general atomic coordinates are given in
table~\ref{Table}. The exact data in table~\ref{Data}. In table~\ref{latpar} the
lattice parameters of the studied compounds are shown.

\begin{table}
  \centering
  \caption{General atomic positions of the ambient temperature phase of hexagonal AMnO$_3$.
$\delta$ indicates a shift of the order of 0.02 lattice parameter.}
  \begin{tabular}{c l l l l c}
    &atom & x & y & z &\\
    &A(1) & 0 & 0 & $\frac{1}{4}$-$\delta$&\\
    &A(2) & $\frac{1}{3}$ & $\frac{1}{3}$ & $\frac{1}{4}$+$\delta$ &\\
    &Mn & $\sim\frac{1}{3}$ & 0 & $\sim 0$ &\\
    &O1 & $\frac{1}{3}+\delta$ & 0 & $\sim\frac{1}{6}$ &\\
    &O2 & $\frac{1}{3}-\delta$ & 0 & $-\sim\frac{1}{6}$ &\\
    &O3 & 0 & 0 & $-\delta$ &\\
    &O4 & $\frac{1}{3}$ & $\frac{1}{3}$ & $\delta$ &\\
  \end{tabular}
  \label{Table}
\end{table}

\begin{table}
  \centering
  \caption{Values for the refinable position in AMnO$_3$. The z-coordinate of Mn is fixed
at zero.}
  \begin{tabular}{c c r r r r}
    &atom & Y & Er & Yb & Lu\\
    &A(1)-z &0.2743& 0.2746 & 0.2753 & 0.2746\\
    &A(2)-z &0.2335& 0.2320 & 0.2326 & 0.2311\\
    &Mn-x & 0.3352 & 0.3396 & 0.3333 & 0.3355\\
    &O1-x & 0.3083 & 0.3113 & 0.3030 & 0.3070\\
    &O1-z & 0.1627 & 0.1645 & 0.1636 & 0.1650\\
    &O2-x & 0.3587 & 0.3593 & 0.3610 & 0.3614\\
    &O2-z &-0.1628 &-0.1620 &-0.1639 &-0.1630\\
    &O3-z &-0.0218 &-0.0225 &-0.0249 &-0.0277\\
    &O4-z & 0.0186 & 0.0186 & 0.0211 & 0.0198\\
  \end{tabular}
  \label{Data}
\end{table}

\begin{table}
  \centering
  \caption{Lattice parameters and unit cell volume for AMnO$_3$.}
  \begin{tabular}{c l r r r r}
    &                 & Y      &     Er &     Yb & Lu\\
    &$a$ (\AA)       & 6.1387 & 6.1121 & 6.0584 & 6.0380 \\
    &$c$ (\AA)       & 11.4071& 11.4200& 11.3561& 11.3610\\
    &$V_{mol}$ (\AA$^3$) & 372.27 & 369.47 & 360.97 & 358.70\\
    \end{tabular}
  \label{latpar}
\end{table}

\section{Detailed structure analysis}

\begin{figure}[htb]
   \centering
   \includegraphics[height=57mm,width=81.5mm]{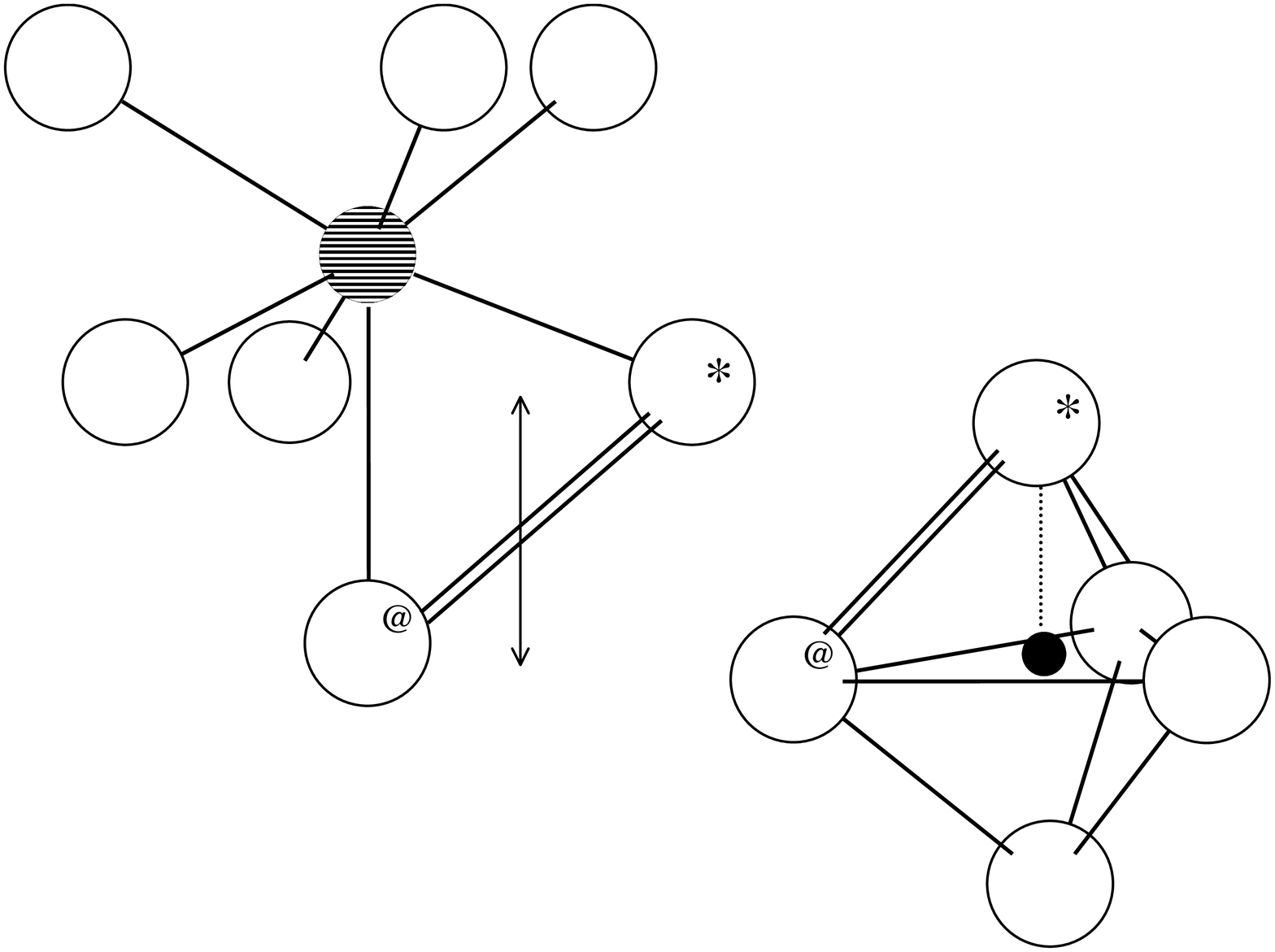}
   \caption{Sketch view of the local environment, showing AO$_7$, left side, and MnO$_5$,
right side. The arrow indicates the distance between two oxygen planes. The
dashed line indicates the Mn-O$_\textrm{Mn,ap}$ distance. Atoms marked with "*"
and with "@" are identical, the double line indicates the shared edge.}
   \label{poly}
\end{figure}

In section~\ref{crystalstructure} we regarded the structure as layers of corner
linked MnO$_5$ trigonal bipyramids, with the A ions between the layers. In
section~\ref{HT vs. LT} the ferroelectric properties were explained using AO$_7$
polyhedra. In this section, the link between the two approaches will be made.
The two polyhedra are shown in Fig.\ref{poly}.

The high temperature phase of the hexagonal AMnO$_3$ consists of 8-fold
coordinated A ions in bicapped antiprisms. The stacking of the AO$_6$ antiprism
is identical to that of three cubic close packed [111] layers, \textsf{ABC}. The
'bicapping' O$_\textrm{A,ap}$ are again located on a \textsf{B} layer, both
above \textsf{C} and below \textsf{A}. In Fig.~\ref{layer} the \textsf{ABC}
stacking can be seen by going from the O$_\textrm{Mn,ap}$ of the bottom layer,
via the A ion, to the O$_\textrm{Mn,ap}$ of the top layer. Note that the
in-plane bond lengths between the O ions and between the A ions is too large to
speak of a true close packed system. In this notation one A ion environment
makes up a \textsf{BABCB} stack, or in chemical elements: OOAOO. The top
O$_{A,ap}$ and the bottom O$_{A,ap}$ of the next layer are one and the same.
Therefore, each B is also the end-member of the next stack, but with the
opposite order \emph{i.e.} \textsf{BCBAB},  in elements: OOAOO. This yields
\textsf{BCBABABCBCBABABCBCBA}, in elements:OOAOOOAOOOAOOOAOOOAO.

Conventionally, the holes between two close packed layers are tetrahedral sites,
\emph{cf.} the spinel structure. In our loosely packed layers, two tetrahedra
from adjacent layers join to form a trigonal bipyramidal site. In AMnO$_3$ this
occurs in the "OOO" stack. Half of the bipyramidal holes are occupied by Mn. In
Fig.~\ref{poly} two neighbouring polyhedra are sketched, where the shared edge
is shown. The O$_\textrm{Mn,ap}$ ions are identical with the oxygens that make
up the antiprisms. The Mn-O$_\textrm{Mn,ap}$ distance is thus equal to the
distance between the antiprism oxygen layer and the O$_\textrm{A,ap}$ layer. The
Mn-O$_\textrm{Mn,ap}$ distance is minimal. Therefore, the steric hindrance of
the Mn restricts this layer separation and increases the A-O$_\textrm{ap}$ bond
length. Thus, the eightfold co-ordination is not uniform. The two
A-O$_\textrm{A,ap}$ have slightly larger bond lengths, \emph{i.e.} $\sim2.7$\AA.

We conclude that hexagonal AMnO$_3$ consists of MnO$_5$ trigonal bipyramids,
stacked in layers, alternated with layers of A ions. This yields a capped
trigonal antiprism as coordination polyhedra for the A ion. In the ferroelectric
phase, the capping is effectively single. This asymmetric coordination of the A
ion yields the ferroelectric properties.

This work is supported by the Netherlands Foundation for the Fundamental
Research on Matter (FOM).

\end{document}